\documentclass[a4paper]{jpconf}
\usepackage{graphicx}
\begin{document}
\title{Shell-model calculations in $^{132}$Sn and $^{208}$Pb regions with 
low-momentum interactions}

\author{A Gargano$^1$, L Coraggio$^1$, A Covello$^{1,2}$ and  N Itaco$^{1,2}$}

\address{$^1$Istituto Nazionale di Fisica Nucleare,\\
Complesso Universitario di Monte S. Angelo, I-80126 Napoli\\
$^2$Dipartimento di Scienze Fisiche, Universit\`{a} di Napoli Federico II,\\
Complesso Universitario di Monte S. Angelo, I-80126 Napoli}

\ead{gargano@na.infn.it}

\begin{abstract}

We discuss shell-model calculations based on the use of low-momentum  interactions derived 
from the free-space nucleon-nucleon potential. A main feature of this approach is the 
construction of a smooth potential, $V_{\rm low-k}$, defined within a given momentum cutoff.
As a practical application of the theoretical framework, we present some selected results of
our current study of nuclei around doubly magic $^{132}$Sn and $^{208}$Pb which have been
obtained starting from the CD-Bonn potential. Focusing attention on the similarity between
the spectroscopy of these two regions, we show that it emerges quite naturally from our
effective interactions without use of any adjustable parameter. 

\end{abstract}

\section{Introduction}
In the last decade, shell-model calculations employing realistic effective interactions 
derived from modern nucleon-nucleon ($NN$) potentials  have entered the main stream of 
nuclear structure theory 
\cite{Coraggio09}. As is well known, the first problem one is confronted with in this kind 
of calculations is the strong short-range repulsion contained in the bare $NN$ potential 
$V_{NN}$, which prevents its direct use in the derivation of the shell-model effective 
interaction $V_{\rm eff}$. The most popular way to overcome this difficulty has long been 
the Brueckner $G$-matrix method. However, a few years ago a new approach  \cite{Bogner02} 
was proposed which consists in deriving from $V_{NN}$ a low-momentum potential, 
$V_{\rm low-k}$, that preserves the deuteron binding energy and scattering phase shifts of
$V_{NN}$ up to a certain cutoff momentum
$\Lambda$. This is a smooth potential which can be used directly to derive $V_{\rm eff}$, 
and it has been shown \cite{Bogner02,Covello03} that it provides an advantageous alternative
to the use of the $G$ matrix.
In this connection, it should be mentioned that  $V_{\rm low-k}$ potentials are currently 
being used in various nuclear theory contexts, such as the study of few-body systems and 
no-core shell-model calculations \cite{Deltuva08,Bogner08}. 

Making use of the $V_{\rm low-k}$ approach, we have recently 
studied~\cite{Coraggio05,Coraggio06,Covello07,Simpson07} several nuclei 
beyond doubly magic  $^{132}$Sn,  showing that their properties are well 
accounted for by a unique shell-model 
Hamiltonian with single-particle energies taken from experiment and two-body effective 
interaction derived from the CD-Bonn $NN$ potential \cite{Machleidt01}. 

Motivated by the very good results obtained in the $^{132}$Sn region and by the existence 
of a close resemblance~\cite{Blomqvist81,Fornal01,Isakov06,Korgul07} between the 
spectroscopy of this region and that of nuclei around stable $^{208}$Pb, we have found it 
challenging to perform a comparative study of these two regions \cite{Covello09}. In this 
paper we present some results from this study, focusing attention on the proton-proton, 
neutron-neutron and proton-neutron multiplets in the three far-from-stability nuclei 
$^{134}$Te, $^{134}$Sn, $^{134}$Sb and in their counterparts in the $^{208}$Pb region, 
$^{210}$Po, $^{210}$Pb, and $^{210}$Bi. 

We start by giving an outline of the theoretical framework in which our shell-model 
calculations are performed and then present and discuss our results. A short summary is 
given in the last section.

\section{Theoretical framework}

In the framework of the shell model an auxiliary one-body potential $U$ is
introduced in order to break up the nuclear Hamiltonian, written as
the sum of the kinetic term and the $NN$ potential, into a one-body 
component $H_0$, which describes the independent motion of the nucleons, 
and a residual interaction $H_1$. Namely,

\begin{equation}
H= \sum_{i=1}^{A} \frac{p^{2}_{i}}{2m} +
\sum_{i<j} V_{ij} = T+ V_{NN}=(T+U)+( V_{NN}-U)=H_{0}+H_{1}.
\label{eq1}
\end{equation}

A reduced model space is then defined in terms of
the eigenvectors of $H_0$ and the diagonalization of the original Hamiltonian 
$H$ in an infinite Hilbert space is reduced to the solution of an eigenvalue 
problem for an effective Hamiltonian $H_{\rm eff}$ in a finite space.

The Hamiltonian  $H_{\rm eff}$ can be derived by way of the $\hat Q$-box 
folded-diagram  expansion (see Ref.~\cite{Coraggio09}). This implies
as first step the calculation of the so-called $\hat Q$-box, which is made up
of  an infinite collection of irreducible and valence-linked
Goldstone diagrams. Once the $\hat Q$-box has been calculated at a given 
order, the infinite series of the folded diagrams has to be summed up.
From this procedure an effective Hamiltonian is obtained containing both
one- and two-body components. Usually, only the two-body term $V_{\rm eff}$ 
is retained, while  the one-body contributions, representing the 
theoretical single-particle energies, are subtracted and replaced 
with single-particle energies taken from experiment~\cite{Shurpin83}.

However, as mentioned in the Introduction no modern $NN$ potential can be
used in a perturbative nuclear structure calculation, unless 
its strong repulsive core is firstly ``smoothed out''.  
Here, we do not embark on any discussion of how  the
effective  Hamiltonian is derived, and  refer to~\cite{Coraggio09}, where 
a detailed description of the whole procedure 
can be found. Rather, in the following  we focus on the  $V_{\rm low-k}$ 
approach. We first outline the essential steps for the  
derivation of the $V_{\rm low-k}$ based on  the Lee-Suzuki similarity 
transformation method~\cite{Suzuki80}, and then describe its main features.

Let us consider the similarity transformation on the Hamiltonian (\ref {eq1}) 

\begin{equation}
\mathcal{H}=X^{-1} H X,
\label{eq2}
\end{equation}

\noindent
where the operator $X$ is defined in the whole Hilbert space.
We now introduce a  cutoff momentum $\Lambda$ that separates fast and slow 
modes to the end of deriving  from the original $V_{NN}$ a low-momentum
potential satisfying a decoupling condition between the 
low- and high-momentum spaces.

The low-momentum space is specified by
 
\begin{equation}
P=\int d \mbox{\boldmath $p$} \mid \mbox{\boldmath $p$}  \rangle \langle 
\mbox{\boldmath $p$} \mid ~,~~~~p \leq \Lambda 
\label{eq3}
\end{equation} 

\noindent
where $\mbox{\boldmath $p$}$ is the two-nucleon relative momentum, and 
the decoupling equation reads

\begin{equation}
Q \mathcal{H} P=0, 
\label{eq4}
\end{equation}

\noindent
with $Q=1-P$ being  the complementary fast-mode space.
The low-momentum Hamiltonian is then given by

\begin{equation}
H_{\rm low-k}=P \mathcal{H} P , 
\label{eq5}
\end{equation}

\noindent
and it can be easily proved that its eigenvalues are a subset 
of the eigenvalues of the original Hamiltonian.

There are, of course, different choices for  the transformation
operator $X$. We take
\begin{equation}
X=e^{\omega}, 
\label{omegaop}
\end{equation}

\noindent
where the wave operator $\omega$ satisfies the conditions:
\begin{equation}
\omega= Q \omega P, 
\label{omegapro1}
\end{equation}
\begin{equation}
P \omega P= Q \omega Q = P \omega Q =0, 
\label{omegapro2}
\end{equation}

\noindent 
the former implying that
\begin{equation}
X=1+ \omega.
\label{omegaop1}
\end{equation}

\noindent
From Eq.~(\ref{eq5}) the low-momentum potential $V_{\rm low-k}$
can be defined as 
\begin{equation}
V_{\rm low-k} = H_{\rm low-k}- PTP.
\label{vvv1}
\end{equation}
Employing transformation (\ref{omegaop1}),  this equation is written as
\begin{equation}
V_{\rm low-k} =  P V_{NN} P +PV_{NN} Q \omega,
\label{vvv}
\end{equation}
\noindent
while Eq.~(\ref{eq4})  becomes
\begin{equation}
Q V_{NN} P + Q H Q \omega - \omega P H P - \omega P V_{NN} Q \omega = 0.
\label{deceq2} 
\end{equation}

\noindent
The solution of Eq. (\ref{deceq2}) gives the  value of $\omega$ needed to 
obtain  $V_{\rm low-k}$.

This decoupling equation can be solved  by means of the
iterative technique for non-degenerate model
spaces proposed in \cite{Andreozzi96}, which is now sketched.
We define the operators:

\begin{eqnarray}
p(\omega) & = & PHP + PHQ \omega,\\
q(\omega) & = & QHQ - \omega PHQ,
\end{eqnarray}
\noindent
in terms of which one can write

\begin{eqnarray}
x_0  & = & - (QHQ)^{-1} QHP ~~, \nonumber \\
x_1 & = & q (x_0)^{-1} x_0 p (x_0) ~~, \nonumber \\
~& ~.~.~.&~ \nonumber \\
x_n & = & q ( x_0+x_1+...+x_{n-1})^{-1} x_{n-1} p ( x_0+x_1+...+x_{n-1}) ~~.
\end{eqnarray}

\noindent
Once the iterative procedure has converged, $x_{n} \rightarrow 0$,
 the operator $\omega$ is given by

\begin{equation}
\omega_n= \sum^n_{i=0} x_i,
\end{equation}

\noindent
In applying this method, we have employed a momentum-space
discretization procedure making use of an adequate number of Gaussian  mesh
points \cite{Krenciglowa76}.

It is worth mentioning that the equation for $ V_{\rm low-k}$ obtained 
using the similarity transformation of Lee and Suzuki is the same as that
one can derive from the $T$-matrix equivalence approach~\cite{Coraggio09,Bogner02}.
This means that the obtained low-momentum potential not only preserves the
deuteron binding energy given by the original $NN$ potential, but also its 
low-momentum ($\leq \Lambda$) half-on-shell $T$ matrix.  

The above  $V_{\rm low-k}$ is however  not Hermitian, which is 
not convenient for various applications, as for instance its use in the 
derivation  of shell-model effective interactions. This $V_{\rm low-k}$ 
may be transformed by means   
of the familiar Schmidt orthogonalization procedure, which leads to a 
Hermitian  new $V_{\rm low-k}$.
As suggested in~\cite{Andreozzi96}, another transformation, based on the 
Cholesky decomposition 
of a symmetric and positive definite matrix, can be used to this end. 
In fact, the matrix $P(1+\omega ^+ \omega)P$,  being symmetric and  positive 
definite,  admits this decomposition,
\begin{equation}
P (1+\omega ^+ \omega)P= PL L^TP,
\end{equation}

\noindent
where $L$ is a lower triangular matrix and  $L^T$ its transpose. 
Since $L$ is real matrix defined  within the $P$-space, we may write 
our transformation as
\begin{equation} 
Z=L^T ,
\end{equation}

\noindent
and the corresponding Hermitian $V_{\rm low-k}$  is
\begin{equation}
V^{\rm chol}_{\rm low-k}=PL^TP(H_0+V_{\rm low-k})P(L^{-1})^TP -PH_0P.
\end{equation}

In Refs.~\cite{Holt04,Coraggio09}, it has been shown that this Hermitian interaction,
 as well all the family of Hermitian interactions which can 
derived from $V_{\rm low-k}$ by means of different transformations,
preserve the full-on-shell $T$ matrix,
and consequently the phase shifts of the original  $V_{NN}$.

The so-obtained $V_{\rm low-k}$ is a smooth potential that can be used directly within the 
$\hat Q$-box
folded diagram theory to derive the shell-model effective interaction.
Actually, it represents an advantageous alternative to the $G$-matrix approach
owing to the fact that it does not depend either on the energy or on the model 
space. This is 
at variance with the $G$ matrix, which is defined in the nuclear medium. In this 
connection, it is worth mentioning that the merit of the
$V_{\rm low-k}$ within the context of realistic shell-model calculations has been assessed by  
several  studies evidencing that $V_{\rm low-k}$ results are as good, or even slightly better 
than, the $G$-matrix ones \cite{Bogner02,Covello03}.

Finally, it is a remarkable feature of the $V_{\rm low-k}$ approach that
different $NN$ potentials lead to low-momentum potentials, which are quite similar to each 
other \cite{Coraggio09,Bogner03}.

\begin{figure}
\begin{center}
\includegraphics[width=37pc]{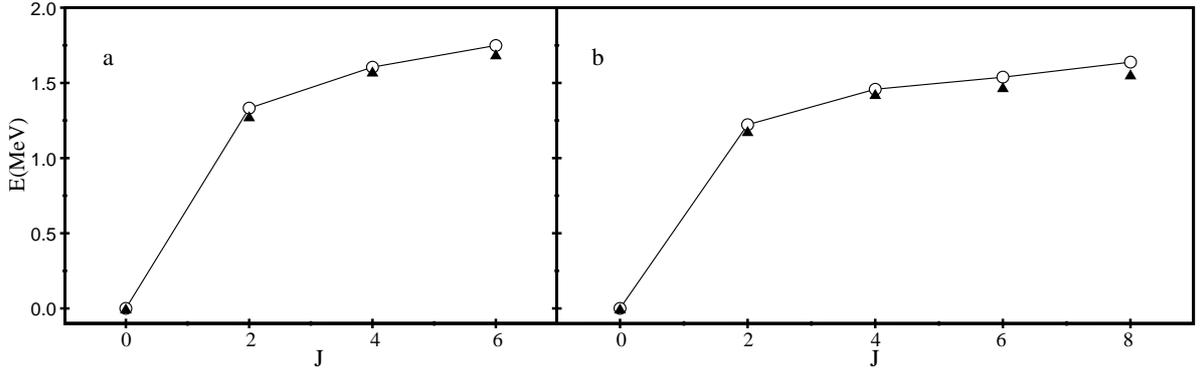}
\end{center}
\caption{\label{fig3}(a) Proton-proton $(\nu f_{7/2})^2$ 
multiplet in $^{134}$Te.(b) Proton-proton $(\nu g_{9/2})^2$ 
multiplet in $^{210}$Po. The theoretical results are represented by open 
circles while the experimental data by solid triangles.}
\end{figure}

\section{Two-valence particle nuclei in $^{132}$Sn and $^{208}$Pb regions}

\subsection{Outline of calculations}
In this paper, we present some results of our current shell-model study of nuclei with two 
valence nucleons in the $^{132}$Sn and $^{208}$Pb regions, which have been obtained 
starting from the CD-Bonn potential renormalized by use of a $V_{\rm low-k}$  with a cutoff 
momentum of $\Lambda=2.2$ fm$^{-1}$. 

In our calculations for $^{134}$Te, $^{134}$Sn, and $^{134}$Sb we assume that the 
the valence protons occupy the five levels 
$0g_{7/2}$, $1d_{5/2}$, $1d_{3/2}$, $2s_{1/2}$, and $0h_{11/2}$ of the
$50-82$ shell, while for the neutrons the model space includes the six levels 
$0h_{9/2}$, $1f_{7/2}$, $1f_{5/2}$, $2p_{3/2}$, $2p_{1/2}$,  and $0i_{13/2}$
of the $82-126$ shell. Similarly, for  $^{210}$Po, $^{210}$Pb, and $^{210}$Bi
we take as model space for the valence protons the six levels of the $82-126$ shell and let 
the valence neutrons occupy the seven levels 
$1g_{9/2}$, $0i_{11/2}$, $0j_{15/2}$, $2d_{5/2}$, $3s_{1/2}$,  $1g_{7/2}$ and $2d_{3/2}$ of 
the $126-184$ shell.
As regards the adopted single-particle neutron and proton energies, they can be found in 
Refs.~\cite{Coraggio05} and  \cite{Coraggio07} for $^{132}$Sn and $^{208}$Pb, respectively.

As mentioned in section~2, the two-body matrix elements of the effective interaction are 
derived within the framework of the $\hat Q$-box folded-diagram expansion. We 
include in the $\hat Q$-box  all diagrams up to second order  in the 
interaction, given by the 
$V_{\rm low-k}$ potential plus the Coulomb force for protons.  
These diagrams are computed within the harmonic-oscillator basis using intermediate states 
composed of all possible hole states and particle states restricted to the five proton and 
neutron shells 
above the Fermi surface. The oscillator parameter is 7.88 MeV for $A=132$ region and 6.88 
MeV for the $A=208$ region, as obtained from the expression 
$\hbar \omega= 45 A^{-1/3} -25 A^{-2/3}$. The calculations have been performed by using the 
NUSHELLX code~\cite{Nushell}.

\subsection{Results}

In figures 1, 2, and 3, we present the experimental~\cite{NNDC,Shergur05} and calculated excitation
energies of the lowest states in the nuclei with  two-proton, two-neutron, and one proton-one neutron
beyond doubly magic $^{132}$Sn and $^{208}$Pb. We see that the agreement between theory 
and experiment is very good for all six nuclei considered, the discrepancies being
well below 100~keV for most of the states.

All the calculated states reported in these figures are dominated by a single configuration,
whose percentage ranges from 80\% to 100\%.   In particular,  they correspond to the  
proton-proton multiplets $(\pi g_{7/2})^2$ and  $(\pi h_{9/2})^2$ in $^{134}$Te and
$^{210}$Po, to the neutron-neutron multiplets $(\nu f_{7/2})^2$ and
$(\nu g_{9/2})^2$ in  $^{134}$Sn and  $^{210}$Pb, and to the proton-neutron multiplets     
$\pi g_{7/2} \nu f_{7/2}$ and  $\pi h_{9/2} \nu g_{9/2}$ in  $^{134}$Sb and $^{210}$Bi.

These  figures evidence the striking resemblance between the behavior of the 
multiplets in the three pairs of counterpart nuclei, $^{134}$Te and $^{210}$Po, $^{134}$Sn and  
$^{210}$Pb, $^{134}$Sb and $^{210}$Bi. We may only note that the curves 
relative to the  $^{208}$Pb neighbors are generally located slightly below those for the counterpart nuclei
in $^{132}$Sn region. 
In particular, from figures~3a and 3b  we see that the two proton-neutron multiplets  show a sizable energy 
gap between the $2^-$ state and the nearly degenerate $0^-$ and $1^-$ states,  as well as a 
distinctive staggering, with the same magnitude and phase,
between the odd and the even members starting from the $3^-$ state.
As regards the two-identical-particle multiplets, figures~1 and 2 show four curves having all the same 
shape. It is worth noting, however, that the curves for  the two-valence-proton 
nuclei are located in an energy interval larger than that pertaining to  the 
two-valence-neutron nuclei. This means that in both
$^{132}$Sn and $^{208}$Pb regions a weakening of the pairing  gap exists for nuclei with
two-valence neutrons  with respect to those with two-valence protons. 

It is worth mentioning that the resemblance between $^{132}$Sn and $^{208}$Pb regions was 
first pointed out by Blomqvist\cite{Blomqvist81}, who noticed that every $^{132}$Sn single-proton and -neutron 
level, characterized by quantum numbers $(n l j)$, has its counterpart around $^{208}$Pb
with quantum numbers $(n l+1 j+1)$. However, until recent years the scarcity of information for
nuclei around $^{132}$Sn, which lies well away from the stability line,
has prevented a detailed comparative study of the two regions.
Nowadays, new data have become available which support the similarity between their 
spectroscopies.  In our calculation, this similarity emerges quite naturally from our effective
interaction  which we have derived from a realistic $NN$ potential.

\begin{figure}
\begin{center}
\includegraphics[width=37pc]{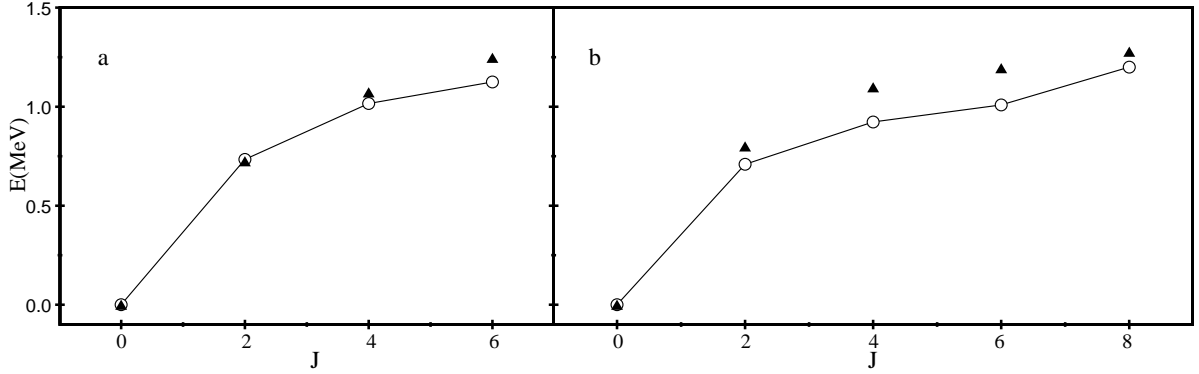}
\end{center}
\caption{\label{fig2}(a) Neutron-neutron $(\pi g_{7/2})^2 $ 
multiplet in $^{134}$Sn.(b) Neutron-neutron $(\pi h_{9/2})^2$ 
multiplet in $^{210}$Pb. The theoretical results are represented by open 
circles while the experimental data by solid triangles.}
\end{figure}

\begin{figure}[h]
\begin{center}
\includegraphics[width=37pc]{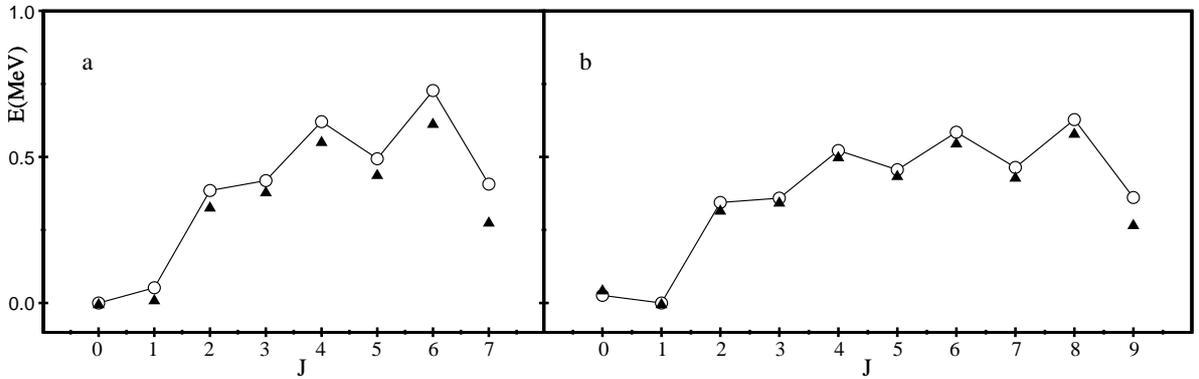}
\end{center}
\caption{\label{fig1}(a) Proton-neutron $\pi g_{7/2} \nu f_{7/2}$ 
multiplet in $^{134}$Sb.(b) Proton-neutron $\pi h_{9/2} \nu g_{9/2}$ 
multiplet in $^{210}$Bi. The theoretical results are represented by open 
circles while the experimental data by solid triangles.}
\end{figure}

\section{Summary}

We  have briefly discussed here the theoretical framework for realistic shell-model 
calculations wherein
use is made of low-momentum interactions derived from the free $NN$ potential. We have 
shown how a smooth low-momentum potential $V_{\rm low-k}$ can be constructed, which 
preserves the deuteron binding energy and scattering phase shifts of the original $V_{NN}$ 
up to a given momentum cutoff. 
We have then presented the results of a shell-model study of nuclei around doubly magic
$^{132}$Sn and $^{208}$Pb, focusing attention on proton-proton, neutron-neutron and 
proton-neutron multiplets. The results obtained for the three nuclei $^{134}$Te, $^{134}$Sn 
and $^{134}$Sb have been compared with those for $^{210}$Po, $^{210}$Pb and $^{210}$Bi, 
which are their counterparts in the region of $^{208}$Pb. In both cases, a low-momentum 
effective interaction derived from the CD-Bonn $NN$ potential has been employed. It should 
be stressed that no adjustable parameter appears in our calculations.

Our results for all six nuclei are in very good agreement with the experimental data and 
account for the striking resemblance between the behavior of the multiplets in the 
$^{132}$Sn and $^{208}$Pb regions. This stimulates further studies to find out whether this 
resemblance extends beyond the two-valence-particle nuclei.

%\medskip

\end{document}